\documentstyle[twoside,fleqn,espcrc2]{article}
\newcommand{\AmS}{{\protect\the\textfont2
  A\kern-.1667em\lower.5ex\hbox{M}\kern-.125emS}}
\newcommand{\vv}{v \cdot v'}

\title{Form Factors for Semileptonic Decays of Heavy-Light Hadrons}

\author{Apoorva Patel\address{CTS and SERC,
        Indian Institute of Science, Bangalore-560012, India \\
        E-mail: adpatel@cts.iisc.ernet.in}}

\begin{document}

\begin{abstract}
The strong coupling lattice QCD solution \cite{APtifr} for the Isgur-Wise
functions, parametrising the semileptonic decay form factors of hadrons
containing a single heavy quark, is reviewed. Several useful features
connected with the result are pointed out.
\end{abstract}

\maketitle

In the quark mass $M\rightarrow\infty$ limit, QCD has an exact $SU(2N_f)$
spin-flavour symmetry. This symmetry relates many matrix elements and form
factors of hadrons containing heavy quarks \cite{Wisgur,NeubPR}. In
particular, the semileptonic weak decay form factors of $s-$wave hadrons
containing a single infinitely heavy quark can be reduced to two unknown
functions, one for mesons ($\xi$) and another for baryons ($\zeta$). The
knowledge of these Isgur-Wise functions is of practical importance in a
precise determination of the quark mixing matrix element $V_{cb}$: the
experimentally measured $B \rightarrow D$ semileptonic decay rates can
be fitted to a leading term involving $\xi$ plus phenomenological estimates
of corrections suppressed by powers of $1/M$.

In the $M\rightarrow\infty$ limit, it is convenient to scale all variables
so as to explicitly separate the factors of heavy quark mass, e.g.
a heavy hadron state is characterised by its four-velocity $v_\mu$ and the
Isgur-Wise functions depend only on the Lorentz invariant $\vv$.
In the static geometry, the light QCD degrees of freedom decouple from the
dynamics of the heavy quark, and flavour independence of QCD fixes the
absolute normalisation of the form factors: $\xi(\vv=1) =1= \zeta(\vv=1)$.

As an explicit case, consider the form factor for semileptonic weak decay
$\overline{B}\rightarrow Dl\overline{\nu}$, and its behaviour as $m_b$ and
$m_c$ go to infinity. Only the partially conserved vector current
contributes:
\begin{eqnarray}
\lefteqn{ \langle D(p') | \bar{c} \gamma_\mu b | \bar{B}(p) \rangle ~=~
  (p+p')_\mu f_+(q^2) } \nonumber \\
  & &+~ {M_B^2 - M_D^2 \over q^2} q_\mu ~[f_0(q^2) - f_+(q^2)] \nonumber \\
  & &{\buildrel M\to\infty \over \longrightarrow}~
  \sqrt{M_B M_D} ~(v+v')_\mu~ \xi(\vv) ~~.
\end{eqnarray}
Here $q=p-p'$ is the momentum transfer, and $\vv = (M_B^2 + M_D^2 - q^2)/
2M_B M_D$. To eliminate the spurious pole at $q^2=0$, we must have $f_0(0)
=f_+(0)$. The divergence of Eq.(1) gives
\begin{eqnarray}
\lefteqn{ {m_b - m_c \over M_B - M_D}
  \langle D(p') | \bar{c} b | \bar{B}(p) \rangle
  ~=~ (M_B + M_D) f_0(q^2) } \nonumber \\
  & &{\buildrel M\to\infty \over \longrightarrow}~
  \sqrt{M_B M_D} ~(1+\vv)~ \xi(\vv) ~~.
\end{eqnarray}

This result contains several noteworthy points: \\
(a) A partially conserved current can undergo an ultraviolet finite
renormalisation. Thus in general a renormalisation constant $Z_0
(M_B,M_D,q^2)$ appears multiplying the matrix element on the l.h.s. of
these equations. According to the Ademollo-Gatto theorem \cite{AdemGat},
the deviation from $1$ of the form factor for the charge operator is of
second order in the symmetry breaking parameter:
\begin{equation}
f_+(M_B \approx M_D,q^2=0) ~=~ 1 + {\cal O}((m_b-m_c)^2) ~~.
\end{equation}
Since only first order deviation from the heavy quark flavour symmetric
limit appears in Eq.(2), it follows that the associated renormalisation
constant obeys $Z_0(M_B=M_D,q^2=0)=1$. In a mass independent
renormalisation scheme, e.g. dimensional regularisation, the
renormalisation constants are independent of the momentum insertion.
In such a case, we can forget about $Z_0$ altogether provided we use
Eq.(2) in the heavy quark flavour symmetric limit only. On the lattice,
with an explicit momentum cutoff, a $q^2$ dependent renormalisation
constant is not ruled out. But the fact that the quark mass operator
is a local point operator on the lattice (note that Fourier transform
of a $\delta-$function is a constant) may still allow one to ``infer''
the continuum result, particularly when the lattice result has a simple
structure. \\
(b) Due to an enhanced kinematic factor on the r.h.s., the quark-hadron
duality analysis of Eq.(2) produces a sum rule yielding a tighter upper
bound for $\xi$ than Bjorken's \cite{BjSLAC}:
\begin{equation}
\xi(\vv) ~\le~ 2/(1+\vv) ~~.
\end{equation}
An implication is that the sum rules containing the scalar operator
would have less contamination from excited states and saturate faster
compared to sum rules containing other operators. \\
(c) The $3-$point function appearing on the l.h.s. obeys a Ward
identity, which relates it to $2-$point functions \cite{toolkit}:
\begin{eqnarray}
\lefteqn{ (m_b-m_c) \langle D(y) | \bar{c} b (q_\mu=0) | B(x) \rangle }
  \nonumber \\
  & &~=~ \langle D(y) | D(x) \rangle ~-~ \langle B(y) | B(x) \rangle ~~.
\end{eqnarray}
This identity guarantees that $Z_0(M_B=M_D,q^2=0)=1$. Moreover, it is
exact even for distinct $m_b$ and $m_c$, and therefore contains enough
information to determine $\xi$ (even though $q_\mu=0$, $\vv$ can be
varied by changing $M_B$ and $M_D$). In practice, the difference of
$2-$point functions may be a simpler object to study than the $3-$point
function, for example while applying QCD sum rules.

In Ref.\cite{APtifr}, $\xi$ was evaluated in strong coupling lattice
QCD, and then the result was converted to continuum language. The
logic of this analysis was inspired by Wilson's renormalisation group
based solution of the Kondo problem \cite{WilRMP}---the weak and strong
coupling results for the same quantity can be related to each other
provided one can evaluate the RG connection between the two with high
precision. It must be kept in mind that the weak and the strong coupling
fixed points describe quite different physics; the scaling relations
which hold at one fixed point may not hold at the other fixed point.
Generically, we can write
\begin{equation}
f(\lambda) = f^* [1 + {\cal O}(a/\lambda)] ~~,~~
  f^* \equiv f(\lambda=\infty) ~~,
\end{equation}
where $\lambda$ is the correlation length. Scaling is exact only on the
critical surface defined by $\lambda=\infty$. The ${\cal O}(a/\lambda)$
terms are non-universal, i.e. they can be different for different
quantities and hence violate scaling. As a consequence, even along the
renormalised trajectory (i.e. for an action containing no lattice
artifacts, yet having a finite cutoff), there exist scaling violations.
For example, Wilson's calculation demonstrates how the spectrum of the
Kondo problem changes under RG evolution. (Another example is the ratio
of two particle binding energy to a single particle mass. It vanishes at
the strong coupling fixed point, while it can be non-zero at the weak
coupling fixed point of an interacting theory.) Thus scaling relations
of the theory must be extracted only in the weak coupling region, after
individually converting each quantity calculated at strong coupling to
its weak coupling analogue.

In QCD scaling violations are ${\cal O}(a\Lambda_{QCD})$, once the
coupling $g^2$ is traded off for a dynamically generated scale. To keep
them under control, it is of paramount importance to select a quantity
which has a simple RG connection between the strong and weak coupling
fixed points, and which still contains the physics of interest. The choice
of the quark mass operator happens to be crucial for this reason---in a
lattice regularisation respecting the quark flavour symmetry, it is a
local point operator with a non-perturbatively constrained renormalisation
constant.

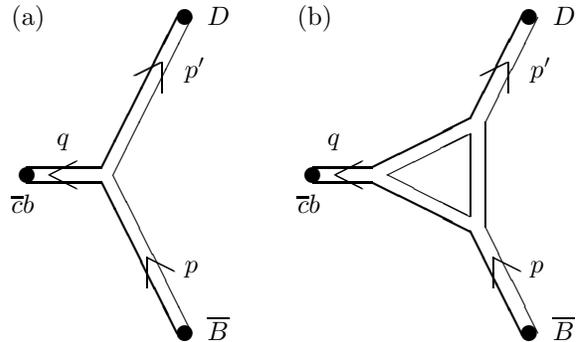
\begin{figure}[t]
{
\setlength{\unitlength}{1mm}
\begin{picture}(75,51)
  \put( 0,44){\makebox(0,0)[bl]{(a)}}
  \put( 2,24){\circle*{2}}
  \put( 0,19){\makebox(0,0)[bl]{$\overline{c}b$}}
  \put(23,45){\circle*{2}}
  \put(26,44){\makebox(0,0)[bl]{$D$}}
  \put(23, 3){\circle*{2}}
  \put(26, 2){\makebox(0,0)[bl]{$\overline{B}$}}
  \thicklines
  \put( 2,25){\line(1, 0){10}}
  \put( 2,23){\line(1, 0){10}}
  \put(12,25){\line(1, 2){10}}
  \put(12,23){\line(1,-2){10}}
  \thinlines
  \put(13.5,24){\line(1, 2){10.5}}
  \put(13.5,24){\line(1,-2){10.5}}
  \put( 6,28){\makebox(0,0)[bl]{$q$}}
  \put( 5,24){\line( 2, 1){3.7}}
  \put( 5,24){\line( 2,-1){3.7}}
  \put(23,37){\makebox(0,0)[bl]{$p'$}}
  \put(20,39){\line( 0,-1){4.1}}
  \put(20,39){\line(-2,-1){3.7}}
  \put(23,11){\makebox(0,0)[bl]{$p$}}
  \put(18,13){\line( 0,-1){4.1}}
  \put(18,13){\line( 2,-1){3.7}}

  \put(38,44){\makebox(0,0)[bl]{(b)}}
  \put(40,24){\circle*{2}}
  \put(38,19){\makebox(0,0)[bl]{$\overline{c}b$}}
  \put(68.8,45){\circle*{2}}
  \put(72,44){\makebox(0,0)[bl]{$D$}}
  \put(68.8, 3){\circle*{2}}
  \put(72, 2){\makebox(0,0)[bl]{$\overline{B}$}}
  \thicklines
  \put(40,25){\line(1, 0){8}}
  \put(40,23){\line(1, 0){8}}
  \put(48,25){\line(2, 1){13}}
  \put(48,23){\line(2,-1){13}}
  \put(61,31.5){\line(1, 2){6.9}}
  \put(61,16.5){\line(1,-2){6.9}}
  \thinlines
  \put(63,31){\line(1, 2){7}}
  \put(63,17){\line(1,-2){7}}
  \put(63,17){\line(0, 1){14}}
  \put(61,18.5){\line(0, 1){11}}
  \put(50,24){\line(2, 1){11}}
  \put(50,24){\line(2,-1){11}}
  \put(42,28){\makebox(0,0)[bl]{$q$}}
  \put(43,24){\line( 2, 1){3.7}}
  \put(43,24){\line( 2,-1){3.7}}
  \put(69,37){\makebox(0,0)[bl]{$p'$}}
  \put(66,39){\line( 0,-1){4.1}}
  \put(66,39){\line(-2,-1){3.7}}
  \put(69,11){\makebox(0,0)[bl]{$p$}}
  \put(64,13){\line( 0,-1){4.1}}
  \put(64,13){\line( 2,-1){3.7}}
\end{picture}
}
\vspace{-13mm}
\caption{Strong coupling diagrams for the scalar form factor in
$\overline{B}\rightarrow Dl\overline{\nu}$ decays: (a) the leading tree
level contribution, (b) an example of the next order loop contribution.}
\label{fig:scdiag}
\end{figure}

The scalar form factor is easily evaluated in the strong coupling
limit, the leading term being a simple pole contribution from the
tree level hadron diagram of Fig.1a. The subleading corrections
arise from diagrams containing light quark loops. The diagrams where
the loops are part of the external $\overline{B}$ and $D$ legs merely
redefine the external parameters. The modification of the form factor
arises from diagrams of type Fig.1b, where the loops interact with
the scalar hadron state. All such diagrams involve at least two more
heavy quark propagators compared to the tree level diagram, and hence
are suppressed in the $M\rightarrow\infty$ limit given the mass
independence of quark-gluon interactions.

Explicitly, with staggered fermions (their chiral symmetries are
essential in restricting the renormalisation of the quark mass operator)
in the strong coupling limit, the scalar form factor is:
\begin{eqnarray}
f_0(q^2) &=& {1 \over 1 - (4/M_{sc}^2) \sum_\alpha \sin^2(q_\alpha/2) }
             \nonumber \\
         &+& {\cal O}(\Lambda_{QCD}^2 / M^2) ~~,
\end{eqnarray}
where $M_{sc}$ is the mass of the scalar meson $\overline{c}b$.
The normalisation of this form factor at $q^2=0$ confirms the expected
non-renormalisation constraint. The ``$\sin$''$-$function in the
denominator is a reflection of the nearest neighbour lattice action
used in the analysis; a more general lattice action would give rise
to a different function. (As a matter of fact the explicit structure
of the strong coupling lattice action generated by RG evolution is
not known for QCD.) What would remain unchanged for any lattice
action, however, is the simple pole structure of $f_0(q^2)$. In the
continuum language, this simple pole corresponds to
\begin{equation}
\xi(\vv) ~=~ 4/(1+\vv)^2 ~~,
\end{equation}
where we have set $M_B=M_D$ and then taken the $M\rightarrow\infty$
limit (which implies $M_{sc}\rightarrow M_B+M_D$). A similar analysis
for the baryon form factor yields $\zeta(\vv)=2/(1+\vv)$.

The result of Eq.(8) fits the experimental data for $\overline{B}\rightarrow
D^*l\overline{\nu}$ decays reasonably well \cite{Argus93,CLEO94}.
It is reassuring to find that the agreement is within $\approx10\%$,
which is the precision of the experimental data as well as the
estimated magnitude (e.g. from QCD sum rule calculations) of the
${\cal O}(1/M)$ difference between the actual form factor and $\xi$.

Encouraged by this agreement, we turn the problem around and construct
a phenomenological model of hadron wavefunctions which reproduces Eq.(8)
exactly. Such wavefunctions can be useful for studying other hadronic
properties. The scalar form factor at $q^2=0$ is nothing but the overlap
of the initial and final state light-cone wavefunctions. Eqs.(2) and (8)
require
\begin{equation}
\int_0^\infty du ~ \phi^*_D(u) ~ \phi_B(u) ~=~
  \Big( {2 \over 1+\vv} \Big)^{3/2}_{q^2=0} ~~,
\end{equation}
where $e^{-u}$ is the fraction of longitudinal light-cone momentum
carried by the heavy quark. A simple choice for the wavefunction is
\begin{equation}
\phi(u) ~=~ 2(M/m_0)^{3/2} ~ u ~ e^{-uM/m_0} ~~,
\end{equation}
with an $M-$independent parameter $m_0$. By replacing the factor of
$u$ with $1-e^{-u}$ and choosing $m_0$ to be two-thirds the constituent
quark mass in the chiral limit, the structure of this wavefunction can
be symmetrised between the two constituent quarks of the meson. It would
be interesting to explore how accurately this wavefunction describes
many other properties of hadrons.


\end{document}